\documentclass{jfm}

\usepackage{graphicx}
\usepackage{newtxtext}
\usepackage{newtxmath}
\usepackage{natbib}
\usepackage{hyperref}

\makeatletter
\let\absfooterflag\@empty
\let\pagelimitfooter\@empty
\def\ps@titlepage{%
  \let\@mkboth\@gobbletwo
  \def\@oddhead{\hfil\thepage}%
  \let\@evenhead\@oddhead
  \let\@oddfoot\@empty
  \let\@evenfoot\@empty}
\patchcmd{\@maketitle}{(Received xx; revised xx; accepted xx)}{}{}{%
  \PackageError{arxiv-preprint}{Could not remove editorial date placeholder}{}}
\makeatother
\usepackage{multirow}
\hypersetup{
    colorlinks = true,
    urlcolor   = blue,
    citecolor  = blue, 
}

\title{The 1/4-power law for bounded variation of wall turbulence from the K\'arm\'an-Howarth equation}

\author{
  Xi Chen\aff{1}\corresp{\email{chenxi97@outlook.com}}
 \and Katepalli R. Sreenivasan\aff{2}\corresp{\email{krs3@nyu.edu}}}

\affiliation{
\aff{1}Key Laboratory of Fluid Mechanics of Ministry of Education, Beihang University (Beijing University of Aeronautics and Astronautics), Beijing, 100191, PR China
\aff{2}Department of Mechanical and Aerospace Engineering, Department of Physics, and the Courant Institute of Mathematical Sciences, New York University, New York, New York 11201, USA
}

\begin{document}
\maketitle

\begin{abstract}

By applying the Kármán–Howarth (KH) equation to turbulent channel flows, we derive the $1/4$-power bounded laws for the peak dissipation and the normalized mean-square streamwise velocity fluctuation. The analysis requires two additional inputs, both standard in the study of turbulent wall flows: the logarithmic mean velocity profile and the near-isotropy of the spatial velocity derivatives. Self-similarity of KH equation yields, for the profile of the mean-square streamwise velocity, the form $c_1 - c_2 (y/\delta)^{1/4}$ in the overlap region \citep{CS2023JFM}, where $\delta$ is the half-channel height; {it also provides the basis for constructing self-similar profiles for high-order moments of streamwise velocity \citep{CS2025PNAS}.} Furthermore, the analysis yields the form $c_1' - c_2' Re^{-1/4}_\tau$ for the inner peak of the streamwise mean-square velocity \citep{CS2022JFM}, and a similar bounded relation for a characteristic wall dissipation \citep{CS2021JFM}.  

\end{abstract}

\section{Introduction}
\label{sec:intro}

In \cite{CS2021JFM,CS2022JFM,CS2023JFM,CS2025PNAS}---referred to below as CS2021, CS2022, CS2023 and CS2025---we developed a framework for describing the scaling of fluctuations in turbulent wall flows. Its central implication is that the Reynolds-number dependence observed in experiments and direct numerical simulations (DNS) is a finite-Reynolds-number effect that vanishes asymptotically. Although the CS papers do not explicitly invoke the attached eddies near the wall \citep{MarusicAndMonty}, the results of those papers do not contradict either their existence or their influence; rather, the framework is intended as a consistent account of the contributions of the eddy structure, whose low-to-moderate Reynolds number manifestation could well be attached eddies. The physical picture considered in CS2021 is that a definite amount of turbulent energy is produced near the wall, a substantial fraction of which is dissipated locally. The remainder, which is transported outwards, sustains turbulence in that region. The outer dynamics, in turn, maintains the turbulent state of the inner region through a complex set of processes discussed extensively in the literature (see, for example, \cite{Hoyasetal2024}). By requiring the smallest outer eddies to account for the imbalance between production and dissipation, we obtained a $-1/4$-power law for the bounded variation of the wall dissipation. This result provided the basis for deriving a number of quantitative scaling relations for velocity and pressure fluctuations in wall flows, which we substantiated through comparisons with experimental and DNS data. In particular, suitable matching conditions enabled us to obtain profiles of pressure and of the mean-square and higher-order velocity fluctuations in the overlap region. Despite the success of these comparisons, the connection between the CS relations and the Navier--Stokes (NS) equations has remained to be established, as has occasionally been pointed out (e.g., \cite{Monkewitz2021}). Establishing this connection is the principal motivation for the present work.

Here, we apply the K\'arm\'an--Howarth (KH) equation, which follows directly from the NS equations, to turbulent channel flow \citep{Maratietal2004} and show how it leads to the principal CS results. We assume the logarithmic variation of the mean velocity profile and the near-isotropy of the spatial velocity derivatives. In particular, we show that self-similarity of the KH equation yields the form $c_1-c_2(y/\delta)^{1/4}$ for the profile of the mean-square streamwise velocity in the overlap region (CS2023), where $\delta$ is the channel half-height. The analysis also yields corresponding power laws for higher-order moments (CS2025). Furthermore, it gives the form $c_1'-c_2'Re_\tau^{-1/4}$ for the inner peak of the mean-square streamwise velocity (CS2022), together with a similar bounded relation for the peak dissipation (CS2021).

Section 2 is a recapitulation of the KH equation as applied to the channel flow, while section 3 discusses the external input in the form of the log law for the mean flow and the Reynolds stress variation near its maximum \citep{Sreenivasan1989, Chenetal2019}. Sections 4 and 5 develop the equations to the first two orders of an asymptotic expansion and combine the results to obtain the principal CS equations. The paper concludes with a summary of results and their implications, and places the present work in a broader context.

\section{K\'arm\'an-Howarth equation for a turbulent channel}
\label{sec:intro2}

We consider the two points $\bf{x}_{+}=\bf{x}+\bf{r}$$/2$ and $\bf{x}_{-}=\bf{x}-\bf{r}$$/2$ in the flow, with their separation distance $\bf{r}$$= (r_1,r_2,r_3)$ and the middle point $\bf{x}$$= (x_1,x_2,x_3)$. Following the Reynolds decomposition, the mean and fluctuation velocities at $\bf{x}_{+}$ are denoted as $U_{i+}=U_i(\bf{x}+\bf{r}$$/2)$ and $u_{i+}=u_i(\bf{x}+\bf{r}$$/2)$, respectively; similarly, $U_{i-}=U_i(\bf{x}-\bf{r}$$/2)$ and $u_{i-}=u_i(\bf{x}-\bf{r}$$/2)$ are the velocities at $\bf{x}_{-}$. Here, $\bf{U}$$=(U_1, U_2,U_3)$ is the mean flow, and $\bf{u}$$=(u_1, u_2,u_3)$ is fluctuation velocity. The difference of fluctuation velocity is defined as $\delta \bf{u}(\bf{x},\bf{r}$$,t) = \bf{u}(\bf{x}+\bf{r}$$/2,t)$$-\bf{u}(\bf{x}-\bf{r}$$/2,t)$, and similarly $\delta \bf{U}(\bf{r}$$) =\bf{U}(\bf{x}+\bf{r}$$/2)$$-\bf{U}(\bf{x}-\bf{r}$$/2)$ for the mean flow. 

By generalizing the KH equation, \cite{Hill2001} has obtained the exact evolution equation for $|\delta {\bf{u}}{|^2}$, for general boundary conditions suitable for incompressible Navier-Stokes equations; see also \cite{Yakhot2001} and \cite{Kurien2001} for this derivation using the Hopf formulation. Adopting the Reynolds decomposition, the KH equation can be written as:
\begin{eqnarray}\label{eq:KWNSfull}
 &&\frac{{\partial \left\langle {|\delta {\bf{u}}|^2} \right\rangle }}{{\partial t}} + \frac{{{U_{i + }} + {U_{i - }}}}{2}\frac{{\partial \left\langle {|\delta {\bf{u}}{|^2}} \right\rangle }}{{\partial {x_i}}} + \frac{{\partial \left\langle {\delta {u_i}|\delta {\bf{u}}{|^2}} \right\rangle }}{{\partial {r_i}}} + \frac{{\partial \left\langle {\delta {U_i}|\delta {\bf{u}}{|^2}} \right\rangle }}{{\partial {r_i}}} \nonumber \\
 &=&  - 2\left\langle {\delta {u_i}\delta {u_j}} \right\rangle \frac{{\partial \delta {U_j}}}{{\partial {r_i}}} - \left\langle {({u_{i + }} + {u_{i - }})\delta {u_j}} \right\rangle \frac{{\partial \delta {U_j}}}{{\partial {x_i}}} - \frac{1}{2}\frac{{\partial \left\langle {({u_{i + }} + {u_{i - }})|\delta {\bf{u}}{|^2}} \right\rangle }}{{\partial {x_i}}} \\\nonumber
 &-& 2\frac{{\partial \left\langle {\delta {u_i}\delta p} \right\rangle }}{{\partial {x_i}}} + \nu \frac{1}{2}\frac{{{\partial ^2}\left\langle {|\delta {\bf{u}}{|^2}} \right\rangle }}{{\partial x_i^2}} + 2\nu \frac{{{\partial ^2}\left\langle {|\delta {\bf{u}}{|^2}} \right\rangle }}{{\partial r_i \partial r_i}} - \left[ {2\nu \left\langle {\frac{{\partial {u_i}}}{{\partial {x _{j + }}}}\frac{{\partial {u_i}}}{{\partial {x _{j + }}}}} \right\rangle  + 2\nu \left\langle {\frac{{\partial {u_i}}}{{\partial {x _{j - }}}}\frac{{\partial {u_i}}}{{\partial {x _{j - }}}}} \right\rangle } \right].
\end{eqnarray}
Here, $v$ is the kinematic viscosity and $p$ is the kinematic pressure fluctuation.

Below, we denote $(u,v,w)$ for $(u_1,u_2,u_3)$, $(U,V,W)$ for $(U_1,U_2,U_3)$, and $(x,y,z)$ for $(x_1,x_2,x_3)$. For the channel, $\langle \rangle$ represents average over the streamwise ($x$) and spanwise ($z$) homogeneous directions, and over time, $t$. For the fully developed state, the mean flow $V=W=0$ while $U(y)$ develops a nontrivial profile in the wall-normal direction. Also, the derivatives in the homogeneous directions for all mean quantities is zero, i.e. $\partial_x\langle \rangle =\partial_z\langle \rangle=0$. Therefore, \eqref{eq:KWNSfull} simplifies to
\begin{eqnarray}\label{eq:KW2}
\frac{{\partial \left\langle {\delta {u_i}|\delta {\bf{u}}{|^2}} \right\rangle }}{{\partial {r_i}}} + \frac{{\partial \left\langle {\delta U|\delta {\bf{u}}{|^2}} \right\rangle }}{{\partial {r_x}}}  + \frac{{\partial \left\langle {{v^\star}|\delta {\bf{u}}{|^2}} \right\rangle }}{{\partial y}} + 2\left\langle {\delta u\delta v} \right\rangle {\left( {\frac{{dU}}{{dy}}} \right)^\star}\nonumber \\
 =  - 2\frac{{\partial \left\langle {\delta v\delta p} \right\rangle }}{{\partial y}} + \frac{\nu }{2}\frac{{{\partial ^2}\left\langle {|\delta {\bf{u}}{|^2}} \right\rangle }}{{\partial {y^2}}} + 2\nu \frac{{{\partial ^2}\left\langle {|\delta {\bf{u}}{|^2}} \right\rangle }}{{\partial {r_i}\partial {r_i}}} - 4{\varepsilon ^\star}.
\end{eqnarray}
Here, the star denotes the mid-point average; that is, $v^\star
=(v_{+}+v_{-})/2$, etc. In $\varepsilon^\star=(\varepsilon_{+}+\varepsilon_{-})/2$, we have $\varepsilon_{\pm}={\nu \left\langle {\frac{{\partial {u_i}}}{{\partial {x _{j \pm }}}}\frac{{\partial {u_i}}}{{\partial {x _{j \pm }}}}} \right\rangle }$, which is the turbulence dissipation rate at $\bf{x}$$_\pm$. 

Consider the specific separation distance between two points $\bf{r}$ $=(r_x,0,r_z)$, which are located on the same wall-parallel plane. It follows immediately that $\delta U=0$, and $\varepsilon^\star=\varepsilon_{+}=\varepsilon_{-}$, 
both of which are equal the dissipation rate $\varepsilon$ anywhere on the plane. For the same reason, $(dU/dy)^\star=dU/dy$. We need to pay some attention to the wall-normal fluctuation velocity, which, through Taylor expansion, can be written $\textrm{v}^{\star} = \textrm{v}+o(r_x, r_z)$. Though this suggests that the star could be omitted for small $r$, it is better to keep it in place for some general $r$.

Moreover, according to definition, the two-point Reynolds stress is $\left\langle {\delta u\delta v} \right\rangle  = 2\left\langle {uv} \right\rangle  - 2\langle \widehat {uv}  \rangle$, where $\widehat {uv} = \left\langle {{u_ + }{v_ - } + {u_ - }{v_ + }} \right\rangle/2$ is the two-point cross term of the Reynolds stress. Hence, \eqref{eq:KW2} is rewritten as 
\begin{eqnarray}\label{eq:prod2}
&&\frac{{\partial \left\langle {\delta {u_i}|\delta {\bf{u}}{|^2}} \right\rangle }}{{\partial {r_i}}} +\frac{{\partial \left\langle {{v^\star}|\delta {\bf{u}}{|^2}} \right\rangle }}{{\partial y}}+ 4\left\langle {u v} \right\rangle { {\frac{{dU}}{{dy}}} }- 4\langle { \widehat {uv}\rangle} { {\frac{{dU}}{{dy}}} } \nonumber \\
& &=-2\frac{{\partial \left\langle {\delta v\delta p} \right\rangle }}{{\partial y}} + \frac{\nu }{2}\frac{{{\partial ^2}\left\langle {|\delta {\bf{u}}{|^2}} \right\rangle }}{{\partial {y^2}}} + 2\nu \frac{{{\partial ^2}\left\langle {|\delta {\bf{u}}{|^2}} \right\rangle }}{{\partial {r_i}\partial {r_i}}} - 4{\varepsilon},
\end{eqnarray}
So far, no assumption has been made. In the limit of $r\to\infty$, \eqref{eq:prod2}
reduces to the exact one-point kinetic energy equation.

Equation \eqref{eq:prod2} describes the energy budget produced by mean shear and Reynolds shear stress, transported and diffused in the $r$ direction (energy cascade) and in the wall-normal direction $y$ (spatial transport), and dissipated by viscous action. The equation was first studied by \cite{Maratietal2004} for turbulent channels, who examined the coexistence of energy cascade and spatial flux effects by applying a further average of \eqref{eq:prod2} over horizontal squares. Later, \cite{Cimarellietal2013} performed a detailed exploration of these dual effects. Recently, by spherical averaging, \cite{Apostolidisetal2023} examined the self-similarity of \eqref{eq:prod2} and obtained Kolmogorov's two-thirds law for the second-order velocity structure function, following the matching procedure of \cite{Lundgren1982}. Furthermore, using atmospheric boundary layer data, \cite{Zhangetal2022} have shown that $\widehat {uv} $ follows the same order of $\langle {uv} \rangle$ and investigated the scaling of third-order velocity structure function in the log region of the mean flow. These studies primarily focus on the Kolmogorov's scaling law. Here, we use the self-similarity of \eqref{eq:prod2} to derive the relations advanced in the CS papers.

\section {Analysis of production term based on the logarithmic mean velocity}
Note that every mean term in \eqref{eq:prod2} depends on $\bf{r}/\delta$, $y/\delta$, and $Re_\tau=u_\tau \delta/\nu $, where $u_\tau$ is wall friction velocity and $\delta$ the half-channel height. To understand the self-similarity, we have to say something about the Reynolds shear stress and the production term.  In the overlap layer given by $1/Re_\tau \ll y/\delta \ll 1$, the mean velocity $U(y)$ satisfies
\begin{equation}\label{eq:log}
\frac{dU}{dy}=\frac{u_\tau}{ \kappa y}
\end{equation}
where the K\'arm\'an constant $\kappa\sim O(1)$, whose precise value is irrelevant to present considerations. Denoting $y^+=y u_\tau/\nu= (y/\delta) Re_\tau$, and substituting \eqref{eq:log} into the following exact mean momentum equation
\begin{equation}\label{eq:MME}
\langle {-u v}\rangle + \nu \frac{dU}{dy}=u^2_\tau (1-\frac{y^+}{Re_\tau}),
\end{equation}
 we have
\begin{equation}\label{eq:uv}
\langle {-u v}\rangle =u^2_\tau (1-\frac{1}{\kappa y^+}-\frac{y^+}{Re_\tau}).
\end{equation}
Therefore, the leading order term (with subscript $L$) of the Reynolds shear stress is
\begin{equation}\label{eq:uvL}
\langle {-u v}\rangle_L=u_\tau^2.
\end{equation}

Note that for increasing $y^+$, the right hand side of \eqref{eq:uv} is a non-monotonic function. Hence, by taking $\partial \langle -uv \rangle/\partial y^+ =0 $, one obtains the $y^+$-location of the maximum Reynolds shear stress from \eqref{eq:uv} to be
\begin{equation}\label{eq:peak}
\frac{1}{\kappa y_p^{+2}}=\frac{1}{Re_\tau},
\end{equation}
so that $y^+_p=(Re_\tau/\kappa)^{1/2}$. Substituting $y^+_p$ into \eqref{eq:uv} yields 
\begin{equation}\label{eq:1uvpeak}
\langle -uv \rangle_p =u_\tau^2 -\frac{u_\tau^2}{\sqrt{\kappa}} Re^{-1/2}_\tau,
\end{equation}
which depicts the deviation of the Reynolds shear stress away from its asymptotic limit $\langle {-u v}\rangle_L$ to be the sub-leading term (with subscript $S$)  
\begin{equation}\label{eq:uvS}
\langle {-u v}\rangle_S=- \frac{u^2_\tau}{\sqrt{\kappa}}  Re^{-1/2}_\tau.
\end{equation}

Multiplication of \eqref{eq:log} by \eqref{eq:1uvpeak} yields the following result:
\begin{equation}\label{eq:prodscale2}
\langle-uv \rangle \frac{dU}{dy}=\langle -uv \rangle_{L} \frac{dU}{dy}+\langle -uv \rangle_{S} \frac{dU}{dy} = \frac{u^3_\tau}{\kappa y} - \frac{u^3_\tau}{\kappa^{3/2} y}  Re^{-1/2}_\tau,
\end{equation}
where 
$\langle- uv \rangle_{L} \frac{dU}{dy} = \frac{u^3_\tau}{\kappa y}$ is the leading order production term, and $\langle- uv \rangle_{S} \frac{dU}{dy} =  -\frac{u^3_\tau}{\kappa^{3/2} y}  Re^{-1/2}_\tau$ is the sub-leading production term.

Since the two-point cross Reynolds stress  $\widehat {uv} = \langle {{u_ + }{v_ - } + {u_ - }{v_ + }} \rangle/2$ is exactly one-point Reynolds shear stress $\langle uv \rangle$  when $r\to 0$, it is reasonable to expect that $\widehat {uv}=\alpha \langle uv \rangle$ with $\alpha = O(1)$. Consequently, we have 
\begin{equation}\label{eq:CrossRS}
\widehat {uv} _L= \alpha \langle uv \rangle_L; \quad \widehat {uv} _S= \alpha \langle uv \rangle_S.
\end{equation}

Another notable point is that the volume integral of production exactly balances dissipation, because of the vanishing spatial transport and scale-to-scale nonlinear transport. In the bulk region of the flow, it is well known that there is a quasi-balance between production and dissipation, which means that $\theta \equiv \varepsilon/(\langle-uv \rangle dU/dy) =O(1)$.  Therefore, as for production, we have 
\begin{equation}\label{eq:dissscale}
\varepsilon_L= \theta \langle-uv \rangle_L \frac{dU}{dy}; \quad  \varepsilon_S= \theta  \langle-uv \rangle_S \frac{dU}{dy}.
\end{equation}
We will use \eqref{eq:prodscale2} and \eqref{eq:dissscale} as external inputs to the analysis in the next section.

\section{Self-similarity analysis of \texorpdfstring{\eqref{eq:prod2}}{(2.3)}}
The above analysis for production and dissipation indicates that the balance of \eqref{eq:prod2} can be expressed to two different orders. Therefore, \eqref{eq:prod2} is decomposed to the leading order balance (with subscript $L$) as
\begin{eqnarray}\label{eq:large}
&&\frac{{\partial {{\left\langle {\delta {u_i}|\delta {\bf{u}}{|^2}} \right\rangle }_L}}}{{\partial {r_i}}} + \frac{{\partial {{\left\langle {{v^ \star }|\delta {\bf{u}}{|^2}} \right\rangle }_L}}}{{\partial y}} + 4{\left\langle {uv} \right\rangle _L}\frac{{dU}}{{d y}} - 4{\langle \widehat {uv}\rangle _L}\frac{{dU}}{{dy}} \nonumber \\
&& =  - 2\frac{{\partial {{\left\langle {\delta v\delta p} \right\rangle }_L}}}{{\partial y}} + \frac{\nu }{2}\frac{{{\partial ^2}{{\left\langle {|\delta {\bf{u}}{|^2}} \right\rangle }_L}}}{{\partial {y^2}}} + 2\nu \frac{{{\partial ^2}{{\left\langle {|\delta {\bf{u}}{|^2}} \right\rangle }_L}}}{{\partial {r_i}\partial {r_i}}} - 4{\varepsilon _L},
\end{eqnarray}
 and the sub-leading order balance (with subscript $S$) as
\begin{eqnarray}\label{eq:small}
&&\frac{{\partial {{\left\langle {\delta {u_i}|\delta {\bf{u}}{|^2}} \right\rangle }_S}}}{{\partial {r_i}}} + \frac{{\partial {{\left\langle {{v^ \star }|\delta {\bf{u}}{|^2}} \right\rangle }_S}}}{{\partial y}} + 4{\left\langle {uv} \right\rangle _S}\frac{{d U}}{{d y}} - 4{\langle \widehat {uv}\rangle _S}\frac{{d U}}{{d y}} \nonumber \\
&& =  - 2\frac{{\partial {{\left\langle {\delta v\delta p} \right\rangle }_S}}}{{\partial y}} + \frac{\nu }{2}\frac{{{\partial ^2}{{\left\langle {|\delta {\bf{u}}{|^2}} \right\rangle }_S}}}{{\partial {y^2}}} + 2\nu \frac{{{\partial ^2}{{\left\langle {|\delta {\bf{u}}{|^2}} \right\rangle }_S}}}{{\partial {r_i}\partial {r_i}}} - 4{\varepsilon _S}.
\end{eqnarray}
To simplify the analysis, we consider the specific case of $r_z=0$, which means that the two points considered lie along the same streamwise line, so that $r_x=r$ and $\partial r_x=\partial r$, with no quantity involving $r_z$ or $r_y$. This, however, does not diminish the generality of the results. One can also have $r_x=r_z=r$ and $\partial r_x= \partial r_z= \partial r$ by regarding the fluctuations as isotropic in the wall-parallel plane. It follows that there is only $r$-dependence in \eqref{eq:large} and \eqref{eq:small}. 

Moreover, while the $y$-derivative for the mean flow is definitely not the $r$-derivative (e.g., $\partial_r U =0\neq \partial _y U$), by the assumption of local isotropy for fluctuating quantities such as $u_i$, $\delta u_i$, we may consider $\partial y=\beta \partial r$, where $\beta=O(1)$. This is because turbulence will be well mixed in the bulk, so the derivatives of fluctuating quantities bear the same relationship to each other as in isotropic turbulence. Indeed, the spherical average of \eqref{eq:prod2}, applied by \cite{Apostolidisetal2023}, supports local isotropy. Consequently, the strain of a fluctuating quantity in the $x_i$-direction could be estimated by the strain in the direction $y$, and hence $r$, especially taking into account the above parameter $\beta$. 

Let us turn attention to other terms in \eqref{eq:large} and \eqref{eq:small}. Denote the leading order velocity 
and length scales for the fluctuation quantity as ($u_L, \ell_L$), and the sub-leading order velocity and length scales as ($u_S, \ell_S$), we may write
\begin{eqnarray}\label{eq:fullf}
\left\langle {\delta {u}|\delta {\bf{u}}{|^2}} \right\rangle (r/\delta,{y^ + },{{\mathop{\rm Re}\nolimits} _\tau }) &=& u_L^3{f_0}\left( {r/{\ell _L},{y^ + }} \right) + u_S^3{f_1}\left( {r/{\ell _S},{y^ + }} \right), \\ \label{eq:fullh}
\left\langle {|\delta {\bf{u}}{|^2}} \right\rangle (r/\delta,{y^ + },{{\mathop{\rm Re}\nolimits} _\tau }) &=& u_L^2{h_0}\left( {r/{\ell _L},{y^ + }} \right) + u_S^2{h_1}\left( {r/{\ell _S},{y^ + }} \right), \\ \label{eq:fullk}
{{\left\langle {{v^ \star }|\delta {\bf{u}}{|^2}} \right\rangle }} (r/\delta,{y^ + },{{\mathop{\rm Re}\nolimits} _\tau }) &=& u_L^3{k_0}\left( {r/{\ell _L},{y^ + }} \right) + u_S^3{k_1}\left( {r/{\ell _S},{y^ + }} \right), \\ \label{eq:fullj}
{{\left\langle {\delta v\delta p} \right\rangle }} (r/\delta,{y^ + },{{\mathop{\rm Re}\nolimits} _\tau }) &=& u_L^3{j_0}\left( {r/{\ell _L},{y^ + }} \right) + u_S^3{j_1}\left( {r/{\ell _S},{y^ + }} \right).
\end{eqnarray}
Here, the large and small velocity scales are of different order, which we may represent as
\begin{equation}\label{eq:usul}
{u_S}/{u_L} = {\mathop{\rm Re}\nolimits} _\tau ^{\gamma} 
\end{equation}
with $\gamma<0$. 
In the following section, we develop a procedure to derive small as well as large velocity and length scales.

\subsection{Leading order analysis}
Substituting the leading order terms of \eqref{eq:fullf}-\eqref{eq:fullj} into \eqref{eq:large}, and using \eqref{eq:log}, we obtain
\begin{eqnarray}\label{eq:solL}
\frac{{u_L^3}}{{{\ell _L}}}\frac{{\partial {f_0}}}{{\partial {r_L}}} + \frac{{u_L^3}}{{{\ell _L}}}\frac{1}{\beta }\frac{{\partial {k_0}}}{{\partial {r_L}}} - \frac{4}{\kappa }\frac{{u_\tau ^3}}{y} + \frac{{4\alpha }}{\kappa }\frac{{u_\tau ^3}}{y} = \nonumber \\ - 2\frac{{u_L^3}}{{{\ell _L}}}\frac{1}{\beta }\frac{{\partial {j_0}}}{{\partial {r_L}}} + \frac{1}{2}\frac{{u_L^2{u_\tau }y}}{{\ell _L^2}}\frac{1}{{{\beta ^2}}}\frac{{(1/{y^ + }){\partial ^2}{h_0}}}{{\partial {r_L}\partial {r_L}}} + 2\frac{{u_L^2{u_\tau }y}}{{\ell _L^2}}\frac{{(1/{y^ + }){\partial ^2}{h_0}}}{{\partial {r_L}\partial {r_L}}} - \frac{4 \theta}{{\kappa}}\frac{{u_\tau ^3}}{y}, 
\end{eqnarray}
{where $r_L=r/\ell_L$.}

To determine the velocity and length scales in \eqref{eq:solL}, we proceed by denoting the unknowns ${u_L} = {u_\tau }{\mathop{\rm Re}\nolimits} _\tau ^A$ and ${\ell _L} = y{\mathop{\rm Re}\nolimits} _\tau ^B$, where $A$ and $B$ have to be determined by the standard similarity requirements. By normalizing \eqref{eq:solL} with $u_\tau^3/y$ and substituting ${u_L} /{u_\tau }={\mathop{\rm Re}\nolimits} _\tau ^A$ and ${\ell _L} /y={\mathop{\rm Re}\nolimits} _\tau ^B$, we have
\begin{eqnarray}
{\mathop{\rm Re}\nolimits} _\tau ^{3A - B}\left(\frac{{\partial {f_0}}}{{\partial {r_L}}} + \frac{1}{\beta}\frac{\partial k_0}{\partial {r_L}}\right)- \frac{4(1-\alpha)}{\kappa } &=& -{\mathop{\rm Re}\nolimits} _\tau ^{3A - B}\left(\frac{2}{\beta}\frac{{\partial {j_0}}}{{\partial {r_L}}} \right) + {\mathop{\rm Re}\nolimits} _\tau ^{2A - 2B}\left(\frac{1}{{2{\beta ^2}}}\frac{{{\partial ^2}{{\tilde h}_0}}}{{\partial {r_L}\partial {r_L}}}\right) \nonumber \\ &+& 2{\mathop{\rm Re}\nolimits} _\tau ^{2A - 2B}\frac{{{\partial ^2}{{\tilde h}_0}}}{{\partial {r_L}\partial {r_L}}} - \frac{4\theta}{{\kappa}}.
\end{eqnarray}
Here, $\tilde h_0(r_L, y^+)=(1/y^+) h_0(r_L, y^+)$ is introduced so that the $(1/y^+)$ factor in the diffusion term of $h_0$ is absorbed by $\tilde h_0$. Furthermore, because $\alpha$, $\beta$, $\theta$ and $\kappa$ are coefficients of $O(1)$, {the $Re$-invariance, or the self-similarity, of $f_0$,$k_0$,$j_0$,$\tilde{h}_0$ with respect to $r_L$ and $y^+$ yields} 
\begin{equation}
{\mathop{\rm Re}\nolimits} _\tau ^{3A - B} \sim O(1) \sim {\mathop{\rm Re}\nolimits} _\tau ^{3A - B}\sim  {\mathop{\rm Re}\nolimits} _\tau ^{2A - 2B}\sim  {\mathop{\rm Re}\nolimits} _\tau ^{2A - 2B}\sim O(1).
\end{equation}
Hence
\begin{equation}
3A-B=2A-2B=0,
\end{equation}
which means that $A=B=0$. Therefore, the scales are actually $u_L=u_\tau$ and $\ell_L=y$. 

So, to the leading order, we have 
\begin{eqnarray}\label{eq:fullfL}
\left\langle {\delta {u}|\delta {\bf{u}}{|^2}} \right\rangle_L (r/\delta,{y^ + },{{\mathop{\rm Re}\nolimits} _\tau }) &=& u_\tau^3{f_0}\left( {r/{y},{y^ + }} \right), \\ \label{eq:fullhL}
\left\langle {|\delta {\bf{u}}{|^2}} \right\rangle_L (r/\delta,{y^ + },{{\mathop{\rm Re}\nolimits} _\tau }) &=& u_\tau^2{h_0}\left( {r/{y},{y^ + }} \right), \\ \label{eq:fullkL}
{{\left\langle {{v^ \star }|\delta {\bf{u}}{|^2}} \right\rangle_L }} (r/\delta,{y^ + },{{\mathop{\rm Re}\nolimits} _\tau }) &=& u_\tau^3{k_0}\left( {r/{y},{y^ + }} \right), \\ \label{eq:fulljL}
{{\left\langle {\delta v\delta p} \right\rangle_L }} (r/\delta,{y^ + },{{\mathop{\rm Re}\nolimits} _\tau }) &=& u_\tau^3{j_0}\left( {r/{y},{y^ + }} \right) .
\end{eqnarray}
The first two expressions have been verified in several previous works, e.g. \cite{Xieetal2021, Apostolidisetal2023}. The third and the forth expressions are new predictions to be verified.

\subsection{Sub-leading order analysis}

Substituting the sub-leading order terms of \eqref{eq:fullf}-\eqref{eq:fullj} into \eqref{eq:small}, using \eqref{eq:log} again, we have
\begin{eqnarray}\label{eq:solS}
\frac{{u_S^3}}{{{\ell _S}}}\frac{{\partial {f_1}}}{{\partial {r_S}}} + \frac{{u_S^3}}{{{\ell _S}}}\frac{1}{\beta }\frac{{\partial {k_1}}}{{\partial {r_S}}} + \frac{4}{\kappa^{3/2} }\frac{{u_\tau ^3}}{y} Re^{-1/2}_\tau - \frac{{4\alpha }}{\kappa^{3/2} }\frac{{u_\tau ^3}}{y} Re^{-1/2}_\tau = \nonumber \\ - 2\frac{{u_L^3}}{{{\ell _S}}}\frac{1}{\beta }\frac{{\partial {j_1}}}{{\partial {r_S}}} + \frac{1}{2}\frac{{u_S^2{u_\tau }y}}{{\ell _S^2}}\frac{1}{{{\beta ^2}}}\frac{{(1/{y^ + }){\partial ^2}{h_1}}}{{\partial {r_S}\partial {r_S}}} + 2\frac{{u_S^2{u_\tau }y}}{{\ell _S^2}}\frac{{(1/{y^ + }){\partial ^2}{h_1}}}{{\partial {r_S}\partial {r_S}}} + \frac{4 \theta}{{\kappa^{3/2}}}\frac{{u_\tau ^3}}{y} Re^{-1/2}_\tau 
\end{eqnarray}
{where $r_S=r/\ell_S$.}

Denoting ${u_S} = {u_\tau }{\mathop{\rm Re}\nolimits} _\tau ^{A'}$ and ${\ell _S} = y{\mathop{\rm Re}\nolimits} _\tau ^{B'}$ and repeating the above calculations, we have
\begin{eqnarray}
{\mathop{\rm Re}\nolimits} _\tau ^{3A' - B'}\left(\frac{{\partial {f_1}}}{{\partial {r_S}}} + \frac{1}{\beta}\frac{\partial k_1}{\partial {r_S}}\right)+ \frac{4(1-\alpha)}{\kappa^{3/2} } Re_\tau^{-1/2} &=& \nonumber \\ -{\mathop{\rm Re}\nolimits} _\tau ^{3A' - B'}\left(\frac{2}{\beta}\frac{{\partial {j_1}}}{{\partial {r_S}}} \right) + {\mathop{\rm Re}\nolimits} _\tau ^{2A' - 2B'}\left(\frac{1}{{2{\beta ^2}}}\frac{{{\partial ^2}{{\tilde h}_1}}}{{\partial {r_S}\partial {r_S}}}\right) + 2{\mathop{\rm Re}\nolimits} _\tau ^{2A' - 2B'}\frac{{{\partial ^2}{{\tilde h}_1}}}{{\partial {r_S}\partial {r_S}}} + \frac{4\theta}{{\kappa^{3/2}}} Re_\tau^{-1/2}
\end{eqnarray}
Here, $\tilde h_1(r_S, y^+)=(1/y^+) h_1(r_S, y^+)$.

Again, because $\alpha$, $\beta$, $\theta$ and $\kappa$ are coefficients of $O(1)$, {the $Re$-invariance of $f_1$,$k_1$,$j_1$,$\tilde{h}_1$ with respect to $r_S$ and $y^+$ yields} 
\begin{equation}
{\mathop{\rm Re}\nolimits} _\tau ^{3A' - B'} \sim Re^{-1/2}_\tau \sim {\mathop{\rm Re}\nolimits} _\tau ^{3A' - B'}\sim  {\mathop{\rm Re}\nolimits} _\tau ^{2A' - 2B'}\sim  {\mathop{\rm Re}\nolimits} _\tau ^{2A' - 2B'}\sim Re^{-1/2}_\tau.
\end{equation}
Hence
\begin{equation}
3A'-B'=2A'-2B'=-1/2,
\end{equation}
and thus $A'=-1/8$ and $B'=1/8$, which means that $u_S=u_\tau Re_\tau^{-1/8}$ and $\ell_S=y Re_\tau^{1/8}$. Noting that $A=0$ and $A'=-1/8$, we have
\begin{equation}
u_S/u_L=u_\tau Re_\tau^{-1/8} /u_\tau =Re_\tau^{-1/8},
\end{equation}
hence $\gamma=-1/8$ (which is indeed negative as one would have expected).

Thus, the sub-leading order terms of the structure functions are 
\begin{eqnarray}\label{eq:fullfS}
\left\langle {\delta {u}|\delta {\bf{u}}{|^2}} \right\rangle_S (r/\delta,{y^ + },{{\mathop{\rm Re}\nolimits} _\tau }) &=& u_\tau^3{f_0}\left( {r/{\ell_S},{y^ + }} \right) Re^{-3/8}_\tau, \\ \label{eq:fullhS}
\left\langle {|\delta {\bf{u}}{|^2}} \right\rangle_S (r/\delta,{y^ + },{{\mathop{\rm Re}\nolimits} _\tau }) &=& u_\tau^2{h_0}\left( {r/{\ell_S},{y^ + }} \right) Re^{-1/4}_\tau, \\ \label{eq:fullkS}
{{\left\langle {{v^ \star }|\delta {\bf{u}}{|^2}} \right\rangle_S }} (r/\delta,{y^ + },{{\mathop{\rm Re}\nolimits} _\tau }) &=& u_\tau^3{k_0}\left( {r/{\ell_S},{y^ + }} \right) Re^{-3/8}_\tau, \\ \label{eq:fulljS}
{{\left\langle {\delta v\delta p} \right\rangle_S }} (r/\delta,{y^ + },{{\mathop{\rm Re}\nolimits} _\tau }) &=& u_\tau^3{j_0}\left( {r/{\ell_S},{y^ + }} \right) Re^{-3/8}_\tau.
\end{eqnarray}

\section{The asymptotic behavior of second order velocity structure function}

Taking the leading order \eqref{eq:fullhL} and the second order \eqref{eq:fullhS} together, we have:
\begin{eqnarray}\label{eq:2ndnew}
\frac{{\left\langle {|\delta {\bf{u}}{|^2}} \right\rangle }}{u_\tau^2} (r/\delta, y^+, Re_\tau)= h_0 \left( r/{\ell_L}, y^+ \right)+ h_1 \left(r/{\ell_S}, y^+ \right) Re^{-1/4}_\tau.
\end{eqnarray}
Other quantities can be put together in similar ways.
Equation \eqref{eq:2ndnew} is a general expression for the longitudinal structure function but we are interested here only in its behavior for large $r$. In this limit, the left hand side will lose the correlation between $\bf{x}_{+}$ and $\bf{x}_{-}$, and becomes independent of $r$, which allows us to write
$h_0(r/r_L\to \infty, y^+)=H_0(y^+)$, and $h_1(r/r_S\to \infty, y^+)=H_1(y^+)$. Equation \eqref{eq:2ndnew} can then be rewritten as
\begin{eqnarray}
\label{eq:2ndnew:r}
\frac{{\left\langle {|\delta {\bf{u}}{|^2}} \right\rangle }}{u_\tau^2} (r\to \infty, y^+, Re_\tau)= H_0 \left( y^+ \right)+ H_1 \left(y^+ \right) Re^{-1/4}_\tau.
\end{eqnarray}
Noting further that ${\left\langle {|\delta {\bf{u}}{|^2}} \right\rangle }= 2\langle u_i u_i \rangle\approx  4\langle u u \rangle $ {(see comment at the end of this paragraph)} for $r\to\infty$, we may write \eqref{eq:2ndnew:r} as
\begin{eqnarray}
\label{eq:uu1a}
\frac{\langle uu \rangle}{u^2_\tau}(y^+, Re_\tau) &\approx & \frac{H_0 (y^+)}{4} + \frac{H_1 (y^+)}{4}  Re^{-1/4}_\tau\\ \label{eq:uu1b}
&=& \frac{H_0 (y^+)}{4} + \frac{H_1 (y^+)}{4 y^{+1/4}}  \left(\frac{y}{\delta}\right)^{1/4}.
\end{eqnarray}
In the bulk region, $\langle uu \rangle/u^2_\tau$ depends only on $y/\delta$ without $y^+$-dependence, so both $H_0$ and  $H_1/y^{+1/4} $ must be constants. Denoting $H_0=4 c_1$ and  $H_1/y^{+1/4} =-4 c_2$, we have 
\begin{eqnarray}\label{eq:3ndnew}
\frac{\langle uu \rangle}{u^2_\tau} \approx c_1 - c_2 \left(\frac{y}{\delta}\right)^{1/4},
\end{eqnarray}
which is the $1/4$-power profile obtained in CS2023. Based on \eqref{eq:3ndnew}, similar $1/4$-power profiles can be obtained for the high-order moments of $\langle uu \rangle$ by a suitable matching procedure, or the linear-$q$-norm-Gaussian process, as explained in CS2025. {Moreover, in writing $\langle u_i u_i \rangle\approx  2\langle u u \rangle$, we have taken $\langle {v}{v} \rangle$ to be small and $\langle ww \rangle$ to be of the same order as $\langle uu \rangle$; indeed, it is known that the latter two have similar profiles for the variance of the spanwise velocity fluctuation, as shown in CS2023.}

Also, \eqref{eq:uu1a} indicates that the gauge function for the inner two-term expansion of equation (2.6) in CS2023 is
\begin{equation}\label{eq:gage}
g(Re_\tau)=Re^{-1/4}_\tau,
\end{equation}
{which immediately leads to the $Re^{-1/4}_\tau$ scaling for the peak value of $\langle uu\rangle$ in CS2022. Alternatively, extending \eqref{eq:uu1a} towards the wall and taking $y^+=15$ in \eqref{eq:uu1a}, one obtains}
\begin{eqnarray}
\label{eq:uu1p}
\frac{\langle uu \rangle }{u^2_\tau}|_{y^+=15} &\approx & \frac{H_0 (y^+=15)}{4} + \frac{H_1 (y^+=15)}{4}  Re^{-1/4}_\tau=c'_1 - c'_2 Re_\tau^{-1/4}.
\end{eqnarray}
This also yields the bounded variation of the peak value with the $-1/4$ power approach to the asymptotic value (CS2022). {Note, however, that \eqref{eq:uu1a} is derived in the overlap region, while we extend it to $y^+=15$ in \eqref{eq:uu1p} because CS2023 has shown that the expansion extends to the peak region as well.}

Finally, given that the gauge function is $Re^{-1/4}_\tau$, it is clear that all peak variables have the same structure as \eqref{eq:uu1p}. In particular, we have the wall dissipation
\begin{equation}\label{eq:diss}
\varepsilon ^+_w = C_1 - C_2 Re^{-1/4}_\tau,
\end{equation}
{($C_1$ and $C_2$ being constants)} which was the basic premise of CS2021. {The extension of \eqref{eq:diss} to the high-order moments of wall dissipation rate is shown in \cite{duanetal2026}.}

\section{Discussion and conclusion}

The scaling theory for the mean velocity in wall-bounded flows was developed by \cite{Prandtl} on the basis of the mixing-length concept and by \cite{Karman} through similarity considerations. These developments led, among other results, to the logarithmic variation of the mean velocity, commonly known as the law of the wall \citep{Coles}. Although the validity and range of applicability of the logarithmic law have occasionally been questioned (e.g., \cite{Barneblatt}), it has emerged as a remarkably good representation of experimental and numerical data (see, for example, \cite{Monkewitz}), reinforcing its status as a cornerstone of the theory of wall-bounded flows.

The behavior of velocity fluctuations, however, presents a rather different picture. Observations at finite Reynolds numbers (see, for example, \cite{Stanford}) show that fluctuations do not obey the scaling that might be expected from a straightforward extension of the law of the wall. In the CS series of papers, we developed a theory showing that the observed Reynolds-number dependence of wall-normalized fluctuations is, in fact, a finite-Reynolds-number effect. More specifically, when normalized by the mean wall shear stress, the peaks of the mean-square fluctuations increase with Reynolds number, but their rate of increase progressively diminishes as the Reynolds number grows. Consequently, the normalized fluctuations approach finite, Reynolds-number-independent limits asymptotically. We refer to this framework as the theory of bounded variations. Comparisons with experimental and DNS data, presented in the CS papers, have been remarkably successful. The theory accounts not only for the peak values but also for the wall-normal distributions of the fluctuations over the entire range of Reynolds numbers currently accessible. It applies to pipe flows and, with appropriate allowance for the effects of outer intermittency, also to boundary layers \citep{Nagib_Marusic_2025}.

The present paper establishes the theoretical basis of the CS framework through the K\'arm\'an--Howarth equation, supplemented by elements of matched asymptotic expansions. It is important to emphasize what this theory does and does not address. It does not seek to establish the existence, structure, or geometry of the eddy structure in wall-bounded flows---in particular, the attached eddies (e.g., \cite{Perry}). Rather, it argues that a model based on the behavior of attached eddies at moderate Reynolds numbers cannot be simply extrapolated to arbitrarily high Reynolds numbers. Our contention is that the present theory provides a consistent description of the fluctuation behavior across the full range of Reynolds numbers, including the asymptotic limit. This view is central to our interpretation of the results and to the context in which the present work should be placed.

\backsection[Acknowledgements]{We appreciate the persistence of the turbulent wall-layer community in asking for a proof of the $1/4$-power bounded law for wall turbulence.}

\backsection[Funding]{
X. Chen's work was supported by the ``Fundamental Research Funds for the Central Universities". KRS's work was supported by an internal grant from New York University.
}

\backsection[Declaration of interests]{ The authors report no conflict of interest.}

\backsection[Data availability statement]{}

\backsection[Author ORCID]{
Xi Chen, http://orcid.org/0000-0002-4702-8735;
Katepalli R. Sreenivasan, https://orcid.org/0000-0002-3943-6827
}




\bibliographystyle{jfm}
\bibliography{jfm}

\end{document}